\documentclass[12pt]{article}
\usepackage{epsfig}
\usepackage{amsfonts}
\usepackage{amscd}
\usepackage{latexsym}
\usepackage{amsmath,amssymb}
\usepackage{verbatim}
\usepackage{setspace}
\usepackage{color}
\usepackage{cite}

\usepackage[textheight=9in, textwidth=6.5in, letterpaper]{geometry}

\numberwithin{equation}{section}

\def\ip{{\mathcal I}^+}
\def\im{{\mathcal I}^-}

\def\e{{\epsilon}}

 \def\p{\partial}

\def\0{{(0)}}
\def\1{{(1)}}
\def\2{{(2)}}
\def\a{{\alpha}}

\def\cO{{\cal O}}

\def\ci{{\mathcal I}}

\def\<{\langle }
\def\>{\rangle }

\newcommand{\bea}{\begin{eqnarray}}
\newcommand{\eea}{\end{eqnarray}}
\newcommand{\be}{\begin{equation}}
\newcommand{\ee}{\end{equation}}
\newcommand{\ba}{\begin{align}}
\newcommand{\ea}{\end{align}}
\def\be{\begin{equation}}
\def\ee{\end{equation}}
\def\beq{\be\begin{array}{c}}
\def\eeq{\end{array}\ee}
\def\Phi_1{E_r }

\newcommand{\w}{z}

\renewcommand{\epsilon}{\varepsilon}

   \makeatletter
  \let\over=\@@over \let\overwithdelims=\@@overwithdelims
  \let\atop=\@@atop \let\atopwithdelims=\@@atopwithdelims
  \let\above=\@@above \let\abovewithdelims=\@@abovewithdelims
\renewcommand\section{\@startsection {section}{1}{\z@}%
                                   {-3.5ex \@plus -1ex \@minus -.2ex}
                                   {2.3ex \@plus.2ex}%
                                   {\normalfont\large\bfseries}}

\renewcommand\subsection{\@startsection{subsection}{2}{\z@}%
                                     {-3.25ex\@plus -1ex \@minus -.2ex}%
                                     {1.5ex \@plus .2ex}%
                                     {\normalfont\bfseries}}

\begin{document}
\begin{titlepage}
\unitlength = 1mm
{\begin{flushright} CALT-TH-2014-164 \end{flushright}}
\begin{center}

\ \\

\vskip 1cm

{ \LARGE {\textsc{Asymptotic Symmetries of  Massless QED  in Even Dimensions}}}

\vspace{0.8cm}
Daniel  Kapec$^\dagger$, Vyacheslav Lysov$^*$ and Andrew Strominger$^\dagger$

\vspace{1cm}

$^\dagger${\it  Center for the Fundamental Laws of Nature, Harvard University,\\
Cambridge, MA 02138, USA}\\

$^*${\it Walter Burke Institute for Theoretical Physics,\\
 California Institute of Technology, \\
 Pasadena, CA 91125, USA}

\begin{abstract}
We consider the scattering of massless particles coupled to an abelian gauge field in $2n$-dimensional Minkowski spacetime. Weinberg's soft photon theorem is recast as Ward identities for infinitely many new nontrivial symmetries of the massless QED $\mathcal{S}$-matrix, with one such identity arising for each propagation direction of the soft photon. These symmetries are identified as large gauge transformations with angle-dependent gauge parameters that are constant along the null generators of null infinity.  Almost all of the symmetries are spontaneously broken in the standard vacuum and the soft photons are the corresponding Goldstone bosons. Our result establishes a relationship between soft theorems and asymptotic symmetry groups in any even dimension. 
\end{abstract}

\vspace{1.0cm}

\end{center}

\end{titlepage}

\pagestyle{empty}
\pagestyle{plain}

\def\vx{{\vec x}}
\def\p{\partial}
\def\po{$\cal P_O$}

\pagenumbering{arabic}

\tableofcontents
\section{Introduction}

Recent work \cite{Strominger:2013lka,Strominger:2013jfa,He:2014laa,Kapec:2014opa,Lysov:2014csa,He:2014cra} has connected long understood soft theorems \cite{Weinberg:1965nx,Low:1958sn,Burnett:1967km,GellMann:1954kc,Gross:1968in} for gauge theory and gravity scattering amplitudes to Ward identities for asymptotic symmetry groups of massless interacting theories coupled to gauge theory and gravity in four dimensions. While several of the soft theorems have been known and understood since the 1960's, many of their associated asymptotic symmetry groups have only recently drawn attention. Conjectures stemming from the correspondence have led to the discovery and investigation of new soft theorems  in four dimensions \cite{Cachazo:2014fwa,Casali:2014xpa,Larkoski:2014hta,  Broedel:2014fsa,Bern:2014vva,White:2014qia,Du:2014eca,Liu:2014vva,Luo:2014wea}, many of which were subsequently identified with asymptotic symmetry groups \cite{Kapec:2014opa,Lysov:2014csa,Campiglia:2014yka}. The leading and subleading soft theorems have been investigated at loop level \cite{Bern:2014oka,He:2014bga,Cachazo:2014dia,Broedel:2014bza}, in the context of string  and (ambi)-twistor string theory \cite{Adamo:2014yya,Geyer:2014lca,Schwab:2014fia,Schwab:2014sla,Bianchi:2014gla}, and have been shown to hold in higher dimensions \cite{Schwab:2014xua,Afkhami-Jeddi:2014fia,Kalousios:2014uva,Zlotnikov:2014sva}.  Some of the asymptotic symmetry groups associated to these new soft theorems, such as the extended BMS group, were previously conjectured \cite{bt,Banks:2003vp}, while the nature of others \cite{Lysov:2014csa} remains unknown. 

Although much work has been done, many questions remain unanswered. The tree-level leading soft-theorem is universal among all theories in arbitrary dimensions. Such striking universality can only be a reflection of an underlying symmetry. 
Motivated by the established correspondence between soft theorems and asymptotic symmetries in four dimensions and the existence of the soft theorems in higher dimensions, we are led to consider the leading soft theorem in massless QED in even-dimensional Minkowski spacetime. The odd-dimensional case is of course also of interest but has additional subtleties which require a separate investigation. We recast the soft theorem in the form of a Ward identity for a new group of asymptotic symmetries. The asymptotic symmetry group in $d=2m+2$ dimensions is the subgroup of the local $U(1)$ large gauge transformations with a gauge parameter given by an unconstrained function on the  $S^{2m}$. Our result generalizes the analysis performed in the four dimensional case \cite{He:2014cra}, further strengthening the relationship between soft theorems and asymptotic symmetries in all even dimensions.

We work in the semiclassical limit and therefore prove the result only at tree-level. However, given that the leading QED soft factor is not renormalized in four dimensions, the result may be exact. Although massless QED is not renormalizable in dimensions greater  than four, we are  
interested in infrared effects where two derivative theories minimally coupled to matter fields serve as good low energy effective theories. The infrared structure of gauge theories is simpler in $d>4$ due to the absence of soft divergences, so we hope that studying the soft theorems and associated asymptotic symmetries in higher dimensions will help to clarify the fate of the new symmetries in the presence of loop corrections both in $d=4$ and $d>4$. 

This paper is organized as follows. In section 2 we review the structure of massless QED in $d=2m+2$ dimensions and establish our coordinates and conventions. In section 3 we restrict our attention to six dimensions for illustrative purposes. We determine appropriate boundary conditions and  introduce the gauge field mode expansions and the relevant soft photon operators. In section 4 we rewrite Weinberg's soft theorem as a Ward identity for local charge operators involving the matter current and the soft photon operators. In section 5 we  demonstrate that these charges generate a new asymptotic symmetry group, which is a subgroup of the original $U(1)$ gauge group. In section 6 we discuss how to generalize our result to the arbitrary even-dimensional case.

 \section{Maxwell's equations in even dimensional Minkowski spacetime}
Abelian gauge theory in $d=2m+2$ dimensional flat Minkowski spacetime is governed by  Maxwell's equations
 \be\label{me}
 \nabla^{\mu}F_{\mu \nu}=e^2J^M_{\nu},  
 \ee
 where $F_{\mu \nu}=\p_{\mu} A_{\nu}-\p_{\nu}A_{\mu}$, $J_{\nu}^M$ is the matter current density, and $e$ is the coupling constant of the theory. The equations (\ref{me}) are  invariant under local gauge transformations of the form 
 \be
 A_{\mu} \to A_{\mu}+\p_{\mu}\epsilon, \hspace{.5 in} \Psi_{\mathcal{Q}} \to e^{i\mathcal{Q}\epsilon}\Psi_{\mathcal{Q}},  
 \ee
 where $\Psi_{\mathcal{Q}}$ is a matter field with electric charge $\mathcal{Q}$. It is useful to introduce retarded coordinates $(u,r,z^a)$ given by 
 \be
x^0 = u+r ,~~~x^i  = r \hat{x}^i(z),  
\ee  
where $u$  is retarded  time and  $\hat{x}^i(z)$ describes  an embedding of the unit  $S^{2m}$ with coordinates $z^a, a=1,..,2m$ into $\mathbb{R}^{2m+1}$ with coordinates $x^i,i=1,..,2m+1$. The flat Minkowski metric then takes the form  
\be
ds^2 = -(dx^0)^2 + (dx^i)^2 = -du^2 -2du dr +r^2 \gamma_{ab} dz^a dz^b .\ee
Here $\gamma_{ab}$ is the metric on the unit radius $S^{2m}$ with covariant derivative $D_a$. In the conformal compactification of Minkowksi spacetime, we can identify future null infinity $(\mathcal{I}^+)$ as the null surface $(r=\infty,u,z^a)$. We also employ advanced coordinates 
\be
x^0 = v-r ,~~~x^i  = -r \hat{x}^i(z),  
\ee  
so that  past null infinity $(\mathcal{I}^-)$ is identified as the surface $(r=\infty,v,z^a)$ and 
\be
ds^2 = -dv^2 +2dv dr +r^2 \gamma_{ab} d\w^a d\w^b.   
\ee
The advanced $S^{2m}$ coordinate $\w$ is antipodally related to the retarded $S^{2m}$ coordinate $z$ in such a way that  null generators of $\mathcal{I}$ passing through spatial infinity are labeled by the same value of $z$ on $\mathcal{I}^+$ and $\mathcal{I}^-$. 
We denote the $u=\pm \infty$ boundaries of $\mathcal{I}^+$ as $\mathcal{I}^+_{\pm}$, and the $v=\pm \infty$ boundaries of $\mathcal{I}^-$ as $\mathcal{I}^-_{\pm}$. 
 Maxwell's equations in retarded coordinates take the form
  \be 
 r^{-2m}\p_r \left(r^{2m} F_{r u}\right)- \p_u F_{r u}+r^{-2}D^a F_{a u}=  e^2J^M_u,  \notag  \ee
\be \label{mxw} -r^{-2m}\p_r \left(r^{2m} F_{ ur}\right)+r^{-2}D^aF_{a r}=  e^2 J^M_r,   \ee
\be r^{-2m+2}\p_r \left(r^{2m-2} \left(F_{r a}-F_{ua}\right)\right)-\p_u F_{r a} +r^{-2} D^b F_{ba}= e^2J^M_a.  \notag
\ee
Similar expressions hold for the advanced coordinates.
The constraint equation for the hypersurface at future null infinity is
\be\label{constraint}
n^\mu \nabla^\nu F_{\nu \mu}=  \frac12 r^{-2m}\p_r \left(r^{2m} F_{r u}\right)- \p_u F_{r u}+r^{-2}D^a \left(F_{a u} -\frac12  F_{a r}\right)= e^2 n^\mu J^M_\mu,  
\ee
where the null normal vector is $n  = \p_u -\frac12 \p_r$.

\section{Six-dimensional  Maxwell primer}
In this section we consider six-dimensional abelian gauge theory at null infinity,\footnote{Janis and Newman studied the null Cauchy problem  for Maxwell's equations in four dimensions in \cite{Janis:1965tx} with similar conclusions.}  postponing the discussion of arbitrary even dimensions to section 6. We determine appropriate boundary conditions for the gauge fields, determine matching conditions to link $\mathcal{I}^-$ quantities to $\mathcal{I}^+$ quantities, and isolate the gauge field zero-mode operators appearing in Weinberg's soft theorem. 
\subsection{Asymptotic analysis at $\mathcal{I}^+$}
We work in retarded radial gauge. The gauge-fixing conditions are 
\be
A_r=0,\;\;\;A_u|_{\ip}=0. \;\;\;    
\ee
This leaves unfixed a residual large gauge symmetry parameterized by an unconstrained function $\epsilon(z)$ on the $S^4$ at $\ip$. Under such a large gauge transformation
\be\label{largegauge}
\delta A_a (z) = \p_a \e(z).  
\ee

In order to analyze the field equations near $\ip$ we assume an asymptotic expansion for the gauge field:
\be
A_a = \sum_{n=0} \frac{A_a^{(n)}}{r^n},\;\;\; A_u = \sum_{n=1} \frac{A_u^{(n)}}{r^n}.  
\ee
The $\cO(r^{-2},r^{-3},r^{-4})$ orders of the constraint equation are (in the absence of matter currents)
\be \label{cnsa} \p_u (A^{(1)}_u -D^a A_a^{(0)})=0,   \ee\be \label{cnsb} 
(D^2-1)A_u^{(1)}+\p_u (2A_u^{(2)}  -D^a A_a^{(1)})=0,   \ee\be \label{cnsc} 
-A_u^{(2)} - \p_u F^{(4)}_{ru} + D^a( F^{(2)}_{a u} -\frac12 F^{(2)}_{ar})=0.  
\ee
In six dimensions, a plane wave has transverse field strength behaving as $F_{ab}\sim {1\over r}$.
Finiteness of the energy flux at each point on  $\ip$ and finiteness of the total energy evaluated on a space-like Cauchy surface
requires
\beq \label{finitefirst} F^{(0)}_{ab}=\p_aA_b^{(0)}-\p_bA_a^{(0)}=0,\\ F^{(0)}_{ub}=\p_u A_b^{(0)}=0,  \eeq
which implies 
\be \label{bc1}
A_a^{(0)} = \p_a \phi(z).  
\ee 
Here $\phi(z)$ is a free, unconstrained function on $S^4$ which will later be identified as the Goldstone mode of the spontaneously broken large gauge symmetry.

The subleading term $A_a^{(1)}(u,z)$ represents the free radiative data.  Finiteness of the total radiated energy requires that at   large values of $|u|$\be\label{bc}
A_a^{(1)}|_{\ip_\pm} =0.  \ee
Finiteness of the Coulombic energy and integration of \eqref{cnsa} then imply 
\be A_u^{(1)}=0.  \ee
Demanding that the electric field fall off like $1 \over r^4$ near $\ci^+_-$ together with  \eqref{cnsb} then imply 
\be \label{bc4} A_u^{(2)}|_{\ip_+}=0,   \ee
and interior values of $A_u^{(2)}$ are  determined by integrating \eqref{cnsb}.
At the next order we must specify the boundary data for the electric field \be  F^{(4)}_{ru}|_{\ip_- }= -3A^{(3)}_{u}|_{\ip_- }\equiv \Phi_1.  \ee  We are interested in scattering processes that revert to the vacuum at $u=\infty$, so we require $F^{(4)}_{ru}|_{\ip_+ }=0$. We additionally require $A_a^{(2)}|_{\ip_{\pm}}=0$.  A full perturbative solution of course requires the equations of motion as well as the constraints.

\subsection{Asymptotic analysis at $\mathcal{I}^-$}
Similar analysis can be applied to a Maxwell field $B_{\mu}$ in advanced coordinates near $\im$. We label the corresponding field strength tensor $G_{\mu\nu}=\p_\mu  B_\nu-\p_\nu B_\mu$, and denote $G_{rv}^{(4)}|_{\mathcal{I}^-_+}=E_r^-$. Advanced radial gauge 
\be
A_r=0,\;\;\;A_v|_{\im}=0
\ee
leaves unfixed a residual large gauge symmetry parameterized by an unconstrained function $\epsilon^-(z)$.  The various finiteness conditions applied in the previous section lead to a similar set of boundary conditions for $B_{\mu}$. In particular, we have  $B^{(0)}_a(z) = \p_a \psi(z)$, with $\psi(z)$ an unconstrained function on $S^4$.

\subsection{Scattering}
Given asymptotic data $A_{\mu}$ on $\mathcal{I}^+$ and $B_{\mu}$ on $\mathcal{I}^-$, we must specify a matching condition for the boundary values of the two gauge fields in order to properly define the scattering problem. In doing so, we also single out a diagonal subgroup of the large gauge transformations acting separately at $\ip$ and $\im$, which can then be interpreted as a symmetry of the $\mathcal{S}$-matrix. 

The boundary condition (\ref{bc}) provides a trivial matching condition for the radiative data.
The only nontrivial component of the gauge field strength on the boundaries of $\ip$ and $\im$ are the quantities $E_r$ and $E_r^-$. As in four dimensions \cite{He:2014cra}, we impose the matching condition
\be
E_r(z)=E_r^-(z).  
\ee
Here $z$ labels a null generator, so that the coordinate argument of $E_r^-$ is antipodally related to the argument of $E_r$.  The corresponding matching condition for the Goldstone modes is 
\be
\phi(z) = \psi(z).  
\ee  
The diagonal subgroup of large gauge transformations acting at $\mathcal{I}^+$ and $\mathcal{I}^-$ is therefore obtained by imposing 
\be \label{match}
\epsilon(z)=\epsilon^-(z).  
\ee

\subsection{Mode expansions}
The radiative modes of the gauge field in the plane wave basis take the form 
\begin{equation}
A_{\mu}(x)=e\sum_{\alpha}\int \frac{d^5q}{(2\pi)^5}\frac{1}{2\omega_q}\left[	\epsilon_{\mu}^{*\alpha}(\vec{q})a_{\alpha}(\vec{q})e^{iqx} + \epsilon_{\mu}^{\alpha}(\vec{q})a_{\alpha}(\vec{q})^{\dagger}e^{-iqx}		\right],  
\end{equation}
where $\omega_q=|\vec{q}\hspace{.02 in} |$,  $\e_\mu^\a$ are the four independent polarization vectors for the photon in six dimensions, and 
\be
[a_\a(\vec{p}), a_\beta(\vec{q})^\dagger ] = 2\omega_q \delta_{\alpha\beta} (2\pi)^5 \delta^5(\vec{p}-\vec{q}).  
\ee
The free radiative data in this basis is of the form 
\begin{equation}
A_a^{(1)}(u,z^a)=-\frac{2\pi^2e}{(2\pi)^5}\p_a \hat{x}^{i}\sum_{\alpha}\int \omega_q d\omega_q [	\epsilon_{i}^{*\alpha}a_{\alpha}(\omega_q \hat{x})e^{-i\omega_q u} 	+\epsilon_{i}^{\alpha}a_{\alpha}(\omega_q \hat{x})^{\dagger}e^{i\omega_q u}].  
\end{equation}
We can define a Fourier image for the radiative modes 
\be\label{ipmodes}
\begin{split}
A_a^{\omega(1)}(z) =  -\frac{e\omega }{8\pi^2}\p_a \hat{x}^{i}(z)\sum_{\alpha} \epsilon_{i}^{*\alpha}a_{\alpha}(\omega \hat{x}(z)),\\
A_a^{-\omega(1)}(z) =  -\frac{e\omega }{8\pi^2}\p_a \hat{x}^{i}(z)\sum_{\alpha} \epsilon_{i}^{\alpha}a_{\alpha}(\omega \hat{x}(z))^\dagger,  
\end{split}
\ee
with $\omega>0$ assumed for both expressions. We can define the corresponding zero mode operator
\be
 A_a^{0(1)} \equiv \frac12 \lim_{\omega \to 0} (A_a^{\omega(1)}+A_a^{-\omega(1)}).  
\ee
In a similar way we can introduce the in-modes 
\be
B^{(1)}_a(v,\w)   \equiv \lim_{r\to \infty} r  \p_a x^i(r,\w) A_i(v-r, r\hat{x}^i(\w))  
\ee
so that 
\be
\begin{split}
B_a^{\omega(1)}(\w) =  -\frac{e\omega }{8\pi^2}\p_a \hat{x}^{i}(\w)\sum_{\alpha} \epsilon_{i}^{*\alpha}a_{\alpha}(-\omega \hat{x}(\w)),\\
B_a^{-\omega(1)}(\w) = - \frac{e\omega }{8\pi^2}\p_a \hat{x}^{i}(\w)\sum_{\alpha} \epsilon_{i}^{\alpha}a_{\alpha}(-\omega \hat{x}(\w))^\dagger.  
\end{split}
\ee
The corresponding zero mode operator is 
\be
 B_a^{0(1)} \equiv \frac12 \lim_{\omega \to 0} (B_a^{\omega(1)}+B_a^{-\omega(1)}).  
\ee

\section{Soft theorem as a Ward identity}
In this section we recast Weinberg's soft theorem as a Ward identity for charges constructed out of the matter and gauge fields. In the following section, we demonstrate that this Ward identity is associated to a new group of asymptotic symmetries of massless QED.  

 \subsection{Soft theorem}
Weinberg's soft theorem takes the same form in any dimension:
 \begin{equation}\label{wst}
\lim_{\omega \to 0}\omega\<z_{n+1},\dots| a_\a(q) \mathcal{S}| z_1,\dots\> =  e\omega\left[ \sum_{k=n+1}^{n+n'} \mathcal{Q}_k\frac{p_k\cdot \epsilon_{\alpha}}{p_k \cdot q}-\sum_{k=1}^{n} \mathcal{Q}_k\frac{p_k\cdot \epsilon_{\alpha}}{p_k \cdot q}    \right] \<z_{n+1},\dots|  \mathcal{S}| z_1,\dots \> .  
 \end{equation}
Here $a_\a(q)$ is a creation operator for an outgoing on-shell photon with polarization $\e_\a$ and momentum $q$. A null momentum vector in 6 dimensions is completely characterized by its energy $\omega$ and a point $z$ on the $S^4$. This allows us to express the soft photon's momenta as
 \begin{equation} \label{q}
 q^{\mu}=\omega \left[1,\hat{x}^i(z)\right].  
 \end{equation}
 Here $\hat{x}(z)$ is the embedding of the unit $S^{4}$  into $\mathbb{R}^{5}$. We use the same parametrization for the momenta of the massless external particles: 
 \begin{equation} \label{p}
 p^\mu_k=E_k\left[1,\hat{x}^i(z_k)\right] .  
 \end{equation}
In states and out states are then determined by the energy $E_k$, electric charge $\mathcal{Q}_k$, and $\ci$-crossing point  $z_k$ for each external particle. We denote the in and out states by
 \be \label{states}
|z_1,\dots,z_n \rangle,\;\;\; \langle z_{n+1},\dots,z_{n+n'}|,  
\ee  
respectively. In what follows, we assume that the incoming and outgoing states do not include soft photons.

Motivated by the expression for the radiative modes (\ref{ipmodes}), we define the function
\begin{equation}
F^{out}_a(z,z_1,\dots,z_{n+n'})\equiv \partial_a\hat{x}^{i}(z)\omega  \sum_{\alpha}  \epsilon^{*\alpha}_{i} \left[     \sum_{k=n+1}^{n+n'} \mathcal{Q}_k\frac{p_k\cdot \epsilon_{\alpha}}{p_k \cdot q}-  \sum_{k=1}^{n} \mathcal{Q}_k\frac{p_k\cdot \epsilon_{\alpha}}{p_k \cdot q}   \right]  
\end{equation}
\begin{equation}
=\sum_{k=n+1}^{n+n'}  \mathcal{Q}_k\p_a\log(1-P(z,z_k))-\sum_{k=1}^{n}  \mathcal{Q}_k\p_a\log(1-P(z,z_k)) .  
\end{equation}
Here we have used the completeness relation for polarization vectors
\begin{equation}
\sum_{\alpha}\epsilon_{\alpha}^{* i}(\vec{q})\epsilon^j_{\alpha}(\vec{q})=\delta^{ij}-\frac{q^iq^j}{\vec{q}^2}   
\end{equation}
and defined a function\footnote{$P$ is known as the invariant distance on the $S^4$, and is related to the cosine of the geodesic distance.}
\be
P(z,z_k)\equiv  \hat{x}_i (z)\hat{x}^i (z_k).  
\ee 
$F^{out}_a(z,z_1,\dots,z_{n+n'})$ (abbreviated $F_a(z;z_k)$) is simply related to the zero mode insertion:
\begin{equation} \label{out}
\langle z_{n+1},\dots|  A_a^{0(1)} (z) \mathcal{S} | z_1,\dots\rangle = -\frac{e^2}{(4\pi)^2}F^{out}_a(z;z_k) \langle z_{n+1},\dots| \mathcal{S} | z_1,\dots\rangle .  
\end{equation}
Straightforward algebra reveals that $F_a^{out}(z;z_k)$ obeys the differential equation
\be \label{diffop}
\sqrt{\gamma}(D^2-2)D^a F^{out}_a(z;z_k) =- (4\pi)^2\left[    \sum_{k=n+1}^{n+n'} \mathcal{Q}_k \;\delta^{4}(z-z_k) -\sum_{k=1}^{n} \mathcal{Q}_k \;\delta^{4}(z-z_k)   \right].  
\ee
We may also consider Weinberg's soft theorem for an incoming soft photon, which reads
 \begin{equation}\label{wst2}
\lim_{\omega \to 0}\omega\<z_{n+1},\dots|  \mathcal{S} a_\a(q)^{\dagger}| z_1,\dots\> = - e\omega \left[\sum_{k=n+1}^{n+n'} \mathcal{Q}_k\frac{p_k\cdot \epsilon_{\alpha}^\ast}{p_k \cdot q} -\sum_{k=1}^{n} \mathcal{Q}_k\frac{p_k\cdot \epsilon_{\alpha}^\ast}{p_k \cdot q} \right]    \<z_{n+1},\dots|  \mathcal{S}| z_1,\dots \> .  
 \end{equation}
We similarly define 
\begin{equation}
F^{in}_a(\w,z_1,\dots,z_{n+n'})\equiv \partial_a\hat{x}^{i}(\w)\omega \sum_{\alpha}\epsilon^{\alpha}_{i}  \left[       \sum_{k=n+1}^{n+n'} \mathcal{Q}_k\frac{p_k\cdot \epsilon^\ast_{\alpha}}{p_k \cdot q} -  \sum_{k=1}^{n} \mathcal{Q}_k\frac{p_k\cdot \epsilon^\ast_{\alpha}}{p_k \cdot q}   \right]   
\end{equation}
\begin{equation}
=-\left[   \sum_{k=n+1}^{n+n'}  \mathcal{Q}_k\p_a\log(1+P(\w,z_k))-\sum_{k=1}^{n}  \mathcal{Q}_k\p_a\log(1+P(\w,z_k)) \right] ,  
\end{equation}
which is in turn related to the zero mode insertion

\begin{equation} \label{in}
\langle z_{n+1},\dots|   \mathcal{S} B_a^{0(1)} (\w)| z_1,\dots\rangle = \frac{e^2}{(4\pi)^2}F^{in}_a(\w;z_k) \langle z_{n+1},\dots| \mathcal{S} | z_1,\dots\rangle .  
\end{equation}
Combining equations (\ref{match}), (\ref{out}), (\ref{diffop}), and (\ref{in}),  we obtain the relation

 \be\label{opward}
 \begin{split}
 \frac1{2e^2}\int  d^4z \sqrt{\gamma}\;\epsilon(z)(D^2-2)D^a \langle z_{n+1},\dots |A_a^{0(1)}(z)\mathcal{S}  |z_1, \dots \rangle   \\
 + \frac1{2e^2}\int  d^4\w \sqrt{\gamma}\;\epsilon^-(\w)(D^2-2)D^a \langle z_{n+1},\dots | \mathcal{S} B_a^{0(1)}(\w) |z_1, \dots \rangle \\
 =\left[    \sum_{k=n+1}^{n+n'}  \mathcal{Q}_k\e(z_k)-\sum_{k=1}^n  \mathcal{Q}_k\e(z_k)   \right]  \langle z_{n+1},\dots |\mathcal{S}|z_1, \dots \rangle.  
 \end{split}
 \ee
We can rewrite this expression as a Ward identity
\be\label{wardid}
\langle z_{n+1},\dots |\left(Q^+_\e \mathcal{S} -\mathcal{S}   Q_\e^-\right)|z_1, \dots \rangle=0,  
\ee
where $Q_\e^\pm$ are charges acting on   $\mathcal{I}^{\pm}$ states. $Q_{\epsilon}^{\pm}$ can be decomposed into a hard charge and a soft charge:
 \be
 Q_\e^\pm = Q_H^\pm + Q_S^{\pm} .  
 \ee
The hard charges  $Q^\pm_H$ are defined so that 
\be \label{qhard}
Q^-_H|z_1,\dots \rangle = \sum_{k=1}^n \mathcal{Q}_k\; \e(z_k)  |z_1,\dots\rangle,\;\;\  \langle z_{n+1},\dots |Q_{H}^+ =   \langle z_{n+1},\dots| \sum_{k=n+1}^{n+n'} \mathcal{Q}_k\; \e(z_k).  
\ee 
The soft charges are given by
\be
\begin{split}
Q^+_S=-\frac1{2e^2}\int d^4z\; \sqrt{\gamma}\;\epsilon(z)\;[D^2-2]D^aA_a^{0(1)}(z),  \\
Q^-_S=\frac1{2e^2}\int d^4\w\; \sqrt{\gamma}\;\epsilon^-(\w)\;[D^2-2]D^aB_a^{0(1)} (\w).  
\end{split}
\ee

\section{From Ward identity to  asymptotic symmetry}
We would now like to interpret the Ward identity (\ref{wardid}) in terms of symmetry transformations on the matter and gauge fields and to identify the asymptotic symmetry group of six dimensional massless QED.
\subsection{Action on matter fields}
Equation (\ref{qhard}) indicates that the charges $Q_{H}^{\pm}$ generate a gauge transformation on the matter fields. We can express $Q_H^{\pm}$ in terms of the gauge current:
\be \begin{split}
Q^+_H=\lim\limits_{r\to \infty}\int_{\mathcal{I}^{+}} r^4 \sqrt{\gamma}dud^4z\;\e(z)\; J^M_u(u,r,z), \\
Q^-_H=\lim\limits_{r\to \infty}\int_{\mathcal{I}^{-}} r^4 \sqrt{\gamma}dvd^4\w\;\e^-(\w)\; J^M_v(v,r,\w).  
\end{split}
\ee 
For a matter field $\Psi_{\mathcal{Q}}$ of charge $\mathcal{Q} $ we have 
\be
[Q^+_H, \Psi_{\mathcal{Q}}(u,r,z) ] =[ \lim_{r\to \infty}\int_{\mathcal{I}^{+}} r^4 \sqrt{\gamma}\;\e J^M_u, \Psi_\mathcal{Q}(u,r,z)] = -\e(z) \mathcal{Q} \Psi_\mathcal{Q}(u,r,z).  
\ee 
The soft charges  $Q_S^{\pm}$ commute with $\Psi_{\mathcal{Q}}$, so 
we see that the total charges $Q_\e^\pm$ generate gauge transformations on the matter fields with gauge parameter $\e(z)$. 

\subsection{Action on gauge fields}
Since the full theory is invariant only under combined gauge transformations of the matter and gauge fields, it is intuitively obvious that $Q^{\pm}_S$ must generate a large gauge transformation for the gauge fields $A_a$ and $B_a$.  In order to make this relationship precise, we can use the constraint equation (\ref{constraint}) along with the boundary conditions from section 3.1 
to rewrite the total charge as a boundary integral 
\be
Q^+_\e =\frac{1}{e^2}\lim_{r\to \infty} \int_{\ip_-} r^4 \sqrt{\gamma}\; d^4z\;\e(z)\; F_{ru}(u,r,z) = \frac{1}{e^2}\int_{\ip_-}  \sqrt{\gamma}\; d^4z\;\e(z)\; E_r(z),  
\ee
\be
Q^-_{\e} =\frac{1}{e^2}\lim_{r\to \infty} \int_{\im_+} r^4 \sqrt{\gamma}\; d^4\w\;\e^-(\w)\; G_{rv}(v,r,\w)= \frac{1}{e^2}\int_{\im_+}  \sqrt{\gamma}\; d^4\w\;\e^-(\w)\; E^-_r(\w).  
\ee

At this point several comments are in order. For $\e(z)=1$ these expressions reduce to the familiar expression for total electric charge at $\ip_-$ and $\im_+$.  For non-constant $\epsilon(z)$ they are the natural generalization of the asymptotic symmetry generators  in the four dimensional case \cite{He:2014cra}.
Both charges are written as  pure boundary integrals of the free data $E_r$ and $E_r^-$, allowing for a canonical identification of asymptotic symmetry transformations at $\mathcal{I}^+$ and $\mathcal{I}^-$.  In the next subsection we demonstrate that it is possible to define a symplectic structure on the phase space of the theory so that the charges do in fact generate large gauge transformations on the gauge fields.

\subsection{Bracket for the free data}
In order to claim that the $Q_\e^{\pm}$ generate gauge transformations we need to define the symplectic structure on the phase space of the theory. The bracket for the radiative modes is unambiguous \cite{Crnkovic:1986ex,Ashtekar:1981bq} and can be deduced from the mode expansion:
\be
[A_a^{(1)}(u,z), \p_{u'} A_b^{(1)}(u',z')] =i \frac{e^2}{2} \gamma_{ab}\; \delta(u-u') \frac{\delta^4(z-z')}{\sqrt{\gamma}}.  
\ee
The bracket for the zero modes can then be defined so that the charge $Q_\epsilon^{\pm}$ generates the correct gauge transformation. The correct bracket is given by
\be\label{cagp}
[E_r(z), \phi(z')] =i e^2 \frac{\delta^4(z-z')}{\sqrt{\gamma}}.  
\ee The bracket (\ref{cagp}) resembles that of the constant modes in \cite{He:2014cra}.
Similar expressions hold for $\im$ quantities. It follows that 
\be
[Q^+_\e, A_a (z)] =i\p_a\e(z),  
\ee
and we conclude that the charges $Q_{\epsilon}^{\pm}$ generate large gauge transformations on the matter fields and gauge fields of the theory. 

As we have seen, $Q_S^{\pm}$ does not annihilate the conventional vacuum of the theory. In fact, when $Q_S^{\pm}$ acts on the vacuum it creates a soft photon, indicating that the large gauge symmetries are spontaneously broken. Under a large gauge transformation with parameter $\epsilon(z)$, the free data $\phi(z)$ transforms as a Goldstone boson: 
\be
\phi(z) \to \phi(z)+\epsilon(z).  
\ee

\section{Generalization to arbitrary even dimensional spacetime}

The results of the preceding sections can be straightforwardly generalized to arbitrary even dimensional flat spacetimes. In this section, we sketch the derivation of the Ward identity for $d=2m+2$ dimensional spacetime, omitting a detailed discussion of the boundary conditions and symplectic form.

The plane wave expansion of the gauge field in $d=2m+2$ dimensions is given by
\begin{equation}\label{2mmode}
A_{\mu}(u,r,z)=e\sum_{\alpha}\int \frac{d^{2m+1}q}{(2\pi)^{2m+1}}\frac{1}{2\omega_q}\left[	\epsilon_{\mu}^{*\alpha}(\vec{q})a_{\alpha}(\vec{q})e^{iqx} + \epsilon_{\mu}^{\alpha}(\vec{q})a_{\alpha}(\vec{q})^{\dagger}e^{-iqx}		\right].  
\end{equation}
Here $\omega_q=|\vec{q}\hspace{.02 in} |$ and  $\a$ labels the $2m$ polarizations of the photon with corresponding polarization vectors $\e_\mu^\a(\vec{q})$.
The operator $a_\a (\vec{q})^\dagger$ is a photon creation operator normalized so that 
\be
[a_\a(\vec{p}), a_\beta(\vec{q})^\dagger ] = 2\omega_q \delta_{\alpha\beta} (2\pi)^{2m+1} \delta^{2m+1}(\vec{p}-\vec{q}).  
\ee
We can evaluate the leading term in the large  $r$ expansion of (\ref{2mmode}) using the saddle point approximation, yielding an expression for the radiative degrees of freedom of the Maxwell field in $d=2m+2$ dimensions near $\mathcal{I}^+$. The expression for the Fourier image  is 
\begin{equation}
A^{\omega(m-1)}_a(z)=\frac{ (-i)^m\omega^{m-1}  e }{2(2\pi)^{m}}\p_a \hat{x}^{j}(z)\sum_{\alpha} 	 \epsilon_{j}^{*\alpha}a_{\alpha}(\omega \hat{x}(z)),    
\end{equation}
\begin{equation}
A^{-\omega(m-1)}_a(z)=\frac{ i^m\omega^{m-1}  e }{2(2\pi)^{m}}\p_a \hat{x}^{j}(z)\sum_{\alpha}	 \epsilon_{j}^{\alpha}a_{\alpha}(\omega\hat{x}(z))^{\dagger},    
\end{equation}
where $\hat{x}^i(z)$ is an embedding of $S^{2m}$ into $\mathbb{R}^{2m+1}$.    We can define a generalized zero mode operator
\begin{equation}
A_a^{0(m-1)}=\frac{1}{2}\lim_{\omega \to 0} (i\omega)^{2-m}\left[	A^{\omega(m-1)}_a+(-1)^mA^{-\omega(m-1)}_a		\right].   
\end{equation}
Using the conventions (\ref{q})-(\ref{states}), we can rewrite Weinberg's soft theorem (\ref{wst}) in the form

\begin{equation} \label{out2}
\langle z_{n+1},\dots|  A^{0(m-1)}_a(z) \mathcal{S} | z_1,\dots\rangle = -\frac{(-1)^me^2}{4(2\pi)^m}F^{out}_a(z;z_k) \langle z_{n+1},\dots| \mathcal{S} | z_1,\dots\rangle .   
\end{equation}
Here the soft factor 
\begin{equation}
F^{out}_a(z,z_1,\dots,z_{n+n'})\equiv \partial_a\hat{x}^{i}(z)\omega  \sum_{\alpha}  \epsilon^{*\alpha}_{i} \left[     \sum_{k=n+1}^{n+n'} \mathcal{Q}_k\frac{p_k\cdot \epsilon_{\alpha}}{p_k \cdot q}-  \sum_{k=1}^{n} \mathcal{Q}_k\frac{p_k\cdot \epsilon_{\alpha}}{p_k \cdot q}   \right]   
\end{equation}
\begin{equation}
=\sum_{k=n+1}^{n+n'}  \mathcal{Q}_k\p_a\log(1-P(z,z_k))-\sum_{k=1}^{n}  \mathcal{Q}_k\p_a\log(1-P(z,z_k))   
\end{equation}
 satisfies the differential equation
\begin{equation}\label{softgen}
\begin{split}
(-1)^{m+1}\sqrt{\gamma}\prod_{l=m+1}^{2m-1}[D^2-(2m-l)(l-1)]D^aF^{out}_a  \\
=\Gamma(m)2^{m}(2\pi)^m\left[   \sum_{k=1}^n \mathcal{Q}_k\delta^{2m}(z-z_k) -\sum_{k=n+1}^{n+n'} \mathcal{Q}_k\delta^{2m}(z-z_k)\right].   
\end{split}
\end{equation}
We can similarly introduce the in-modes
\begin{equation}
B^{\omega(m-1)}_a(z)=\frac{ i^m\omega^{m-1}  e }{2(2\pi)^{m}}\p_a \hat{x}^{j}(z)\sum_{\alpha} 	 \epsilon_{j}^{*\alpha}a_{\alpha}(-\omega \hat{x}(z)),  
\end{equation}
\begin{equation}
B^{-\omega(m-1)}_a(z)=\frac{ (-i)^m\omega^{m-1}  e }{2(2\pi)^{m}}\p_a \hat{x}^{j}(z)\sum_{\alpha}	 \epsilon_{j}^{\alpha}a_{\alpha}(-\omega\hat{x}(z))^{\dagger},
\end{equation}
and the associated zero mode operator
\begin{equation}
B_a^{0(m-1)}=\frac{1}{2}\lim_{\omega \to 0} (i\omega)^{2-m}\left[	B^{\omega(m-1)}_a+(-1)^mB^{-\omega(m-1)}_a		\right].   
\end{equation}
We then have
\begin{equation} \label{outmodes}
\langle z_{n+1},\dots|  \mathcal{S}B^{0(m-1)}_a(z) | z_1,\dots\rangle = \frac{e^2}{4(2\pi)^m}F^{in}_a(z;z_k) \langle z_{n+1},\dots| \mathcal{S} | z_1,\dots\rangle ,   
\end{equation}
where 
\begin{equation}
F^{in}_a(\w,z_1,\dots,z_{n+n'}) 
=-\left[   \sum_{k=n+1}^{n+n'}  \mathcal{Q}_k\p_a\log(1+P(\w,z_k))-\sum_{k=1}^{n}  \mathcal{Q}_k\p_a\log(1+P(\w,z_k)) \right] .   
\end{equation}

Combining equations (\ref{match}), (\ref{out2}), (\ref{softgen}), and (\ref{outmodes}), we can rewrite the soft theorem as the Ward identity
\be\label{wardid2m}
\langle z_{n+1}\dots |\left(Q^+_\e \mathcal{S} -\mathcal{S}   Q_\e^-\right)|z_1, \dots \rangle=0.   
\ee
The charges $Q_\e^\pm= Q_H^\pm + Q_S^{\pm}$  act on   $\mathcal{I}^{\pm}$ states, with the action of $Q_H^{\pm}$ defined so that 
\be \label{qhard2}
Q^-_H|z_1,\dots \rangle = \sum_{k=1}^n \mathcal{Q}_k\; \e(z_k)  |z_1,\dots\rangle,\;\;\  \langle z_{n+1},\dots |Q_{H}^+ =   \langle z_{n+1},\dots| \sum_{k=n+1}^{n+n'} \mathcal{Q}_k\; \e(z_k).   
\ee 
The soft charges take the form 
\be\label{scgen}
Q^+_S =  -\frac1{2e^2} \frac{2^{2-m}} {\Gamma(m)}    \int  d^{2m}z \sqrt{\gamma} \; \e(z)  \prod\limits^{2m-1}_{l=m+1}  (D^2  -(2m-l)(l-1) ) D^a  A^{0(m-1)}_a ,  
\ee
\be\label{scgen}
Q^-_S =  \frac{(-1)^{m}}{2e^2} \frac{2^{2-m}} {\Gamma(m)}     \int  d^{2m}z \sqrt{\gamma} \; \e^-(z)  \prod\limits^{2m-1}_{l=m+1}  (D^2  -(2m-l)(l-1) ) D^a  B^{0(m-1)}_a . 
\ee
Note that the careful limiting procedure of section four may still be applied to $A_a^{\omega}$ near $\omega=0$. The charges $Q_{H}^{\pm}$ can be written in terms of the gauge current 
\be Q^+_H=\lim\limits_{r\to \infty} r^{2m}  \int_{\ip}\sqrt{\gamma}\;\e(z)\; J^M_u(u,r,z)  , \;\;\;\;Q^-_H=\lim\limits_{r\to \infty} r^{2m}  \int_{\im}\sqrt{\gamma}\;\e^-(z)\; J^M_v(v,r,z) .
\ee
This operator generates a gauge transformation with parameter $\e(z)$ when acting on the matter fields. 
Assuming a natural generalization of the boundary conditions from section 3 and using Maxwell's equations (\ref{mxw}), we can write the total charge $Q_\e^\pm = Q_H^{\pm}+Q_S^{\pm}$ as a boundary integral
\be\label{charge_in_d} \begin{split} 
Q_\e^+ =\frac{1}{e^2} \lim_{r\to \infty} r^{2m} \int_{\ip_-} d^{2m}z \;\sqrt{\gamma} \; \e(z)\; F_{ru}(u,r,z),  \\
Q_\e^- =\frac{1}{e^2} \lim_{r\to \infty} r^{2m} \int_{\im_+} d^{2m}z \;\sqrt{\gamma} \; \e^-(z)\; G_{rv}(v,r,z).
\end{split}
\ee
We can introduce an extended phase space for the modes on $\ip$ and $\im$ to include $\phi(z)$ and $E_r(z)$. The symplectic form of section 5.3 can then be used to demonstrate that (\ref{charge_in_d}) generates large gauge transformations on the matter fields and gauge fields.  These large gauge transformations are the asymptotic symmetries for even dimensional  massless QED.

\section*{Acknowledgements}
We are  grateful  to T. He, D. Jafferis, P. Mitra,   S. Pasterski, A. Porfyriadis, M. Schwartz and A. Zhiboedov for useful conversations. This work was supported in part by DOE grant DE-FG02-91ER40654 and the Fundamental Laws Initiative at Harvard.
The work of V.L. is supported in part by the Sherman Fairchild scholarship and DOE grant DE-SC0011632.

\end{document}